\documentclass[12pt]{article}
\usepackage[margin=22mm]{geometry}
\usepackage{graphicx}
\usepackage{amssymb,amsfonts,amsmath}
\usepackage{apalike}
\newcommand{\equn}[1]{\begin{equation}\label{#1}}
\newcommand{\eqan}[1]{\begin{eqnarray}\label{#1}}
\newcommand{\eqa}{\begin{eqnarray}}
\newcommand{\equ}{\begin{equation}}
\newcommand{\nuqe}{\end{equation}}
\newcommand{\uqe}{\end{equation}}
\newcommand{\naqe}{\end{eqnarray}}
\newcommand{\aqe}{\end{eqnarray}}

\newcommand{\la}{\langle}
\newcommand{\ra}{\rangle}
\newcommand{\e}{{\rm e}}

\begin{document}
\title{Colored extrinsic fluctuations and stochastic gene expression}
\author{Vahid Shahrezaei\footnote{Correspondence: vahid@cnd.mcgill.ca; Tel: +1 514 398 8092; Fax: +1 514 398 7452}
\and
Julien F.\ Ollivier
\and
Peter S.\ Swain\footnote{Correspondence: swain@cnd.mcgill.ca; Tel: +1 514 398 4360; Fax: +1 514 398 7452}
}
\date{}
\maketitle
\nocite{TitlesOn}

\vspace*{-13mm}
\begin{center}
\begin{minipage}{15cm}
{\small 
Centre for Non-linear Dynamics, Dept.\ of Physiology, McGill University, 3655 Promenade Sir William Osler, Montreal, Quebec H3G 1Y6, Canada
}
\end{minipage}
\end{center}

\vspace*{5mm}

\subsection*{Abstract}
Stochasticity is both exploited and controlled by cells. Although the intrinsic stochasticity inherent in biochemistry is relatively well understood, cellular variation is predominantly generated by interactions of the system of interest with other stochastic systems in the cell or its environment. Such extrinsic fluctuations are non-specific, affecting many system components, and have a substantial lifetime comparable to the cell cycle (they are `colored'). Here we extend the standard stochastic simulation algorithm to include extrinsic fluctuations. We show that these fluctuations affect mean protein numbers and intrinsic noise, can speed up typical network response times, and explain trends in high-throughput measurements of variation. If extrinsic fluctuations in two components of the network are correlated, they may combine constructively (amplifying each other) or destructively (attenuating each other). Consequently, we predict that incoherent feedforward loops attenuate stochasticity, while coherent feedforwards amplify it. Our results demonstrate that both the timescales of extrinsic fluctuations and their non-specificity substantially affect the function and performance of biochemical networks.

\subsubsection*{Keywords}
biochemical networks / extrinsic noise / intrinsic noise / stochastic simulation algorithm

\subsubsection*{Subject category}
Computational methods / Metabolic and regulatory networks

\subsubsection*{Journal reference}
Molecular Systems Biology 4:196 (6 May 2008). Supplementary information is available at the Molecular Systems Biology website (www.nature.com/msb). 

\newpage
\subsection*{Introduction}
Biochemical networks are stochastic: fluctuations in numbers of molecules are generated intrinsically by the dynamics of the network and extrinsically by interactions of the network with other stochastic systems \cite{elowitz02,swain02}. Stochastic effects in protein numbers can drive developmental decisions \cite{arkin98,suel,maamar,nachman}, be inherited for several generations \cite{rosenfeld05,kaufmann}, and have perhaps influenced the organization of the genome \cite{swain04,becskei05}. Intrinsic fluctuations are generated by thermal fluctuations affecting the timing of individual reactions. Their magnitude is increased by low copy numbers. The source of extrinsic fluctuations, however, is mostly unknown \cite{kaern05}, although cell cycle effects \cite{rosenfeld05,volfson} and upstream networks \cite{volfson} contribute. Yet extrinsic fluctuations dominate cellular variation in both prokaryotes \cite{elowitz02} and eukaryotes \cite{raser}. They are colored, having a lifetime that is not negligible but comparable to the cell cycle \cite{rosenfeld05}, and they are non-specific, potentially affecting equally all molecules in the system \cite{pedraza}. They are thus difficult to model and their effects hard to predict \cite{volfson,sigal,geva,austin,cox06,tsimring,scott,wolde}.

Intrinsic and extrinsic stochasticity can be measured by creating a copy of the network of interest in the same cellular environment as the original network \cite{elowitz02}. We can then define intrinsic and extrinsic variables, and their fluctuations generate intrinsic and extrinsic stochasticity, or noise \cite{swain02}. Intrinsic variables typically specify the copy numbers of the molecular components of the network. Their values differ for each copy of the network. Extrinsic variables often describe molecules that equally affect each copy of the network. Their values are therefore the same for each copy. Considering gene expression, the number of transcribing RNA polymerases is an intrinsic variable (it is different for each copy of the network), whereas the number of cytosolic RNA polymerases is an extrinsic variable (both copies of the network are exposed to the same cytosolic RNA polymerases). 

Noise is quantified by measuring an intrinsic variable, for example, the number of proteins, for both copies of the network. Fluctuations of the intrinsic variable will have intrinsic and extrinsic components: intrinsic variables are themselves part of a stochastic system and that system interacts with other stochastic systems. Throughout we will use the term `noise' to exclusively mean a measure of stochasticity, usually the coefficient of variation. Experimentally, the relative number of proteins can be quantified in living cells using fluorescent proteins \cite{elowitz02,raser,ozbudak02,blake}. Denoting $I_1$ as the intrinsic variable (the number of proteins) for the first copy of the system and $I_2$ the equivalent for the second copy, then intrinsic noise is determined by a measure of the difference between $I_1$ and $I_2$ because intrinsic fluctuations cause variation in $I_1$ to be uncorrelated with that of $I_2$. Extrinsic fluctuations, however, cause variation in $I_1$ and $I_2$ to be correlated because they equally affect both copies of the system. Extrinsic noise is a measure of this correlation and is determined by the cross-correlation function of $I_1$ and $I_2$. The squares of the intrinsic and the extrinsic noise sum to give the square of the total noise of the intrinsic variable, which is defined as its coefficient of variation \cite{swain02}. 

Here we consider the effects of extrinsic fluctuations on biochemical networks. Extrinsic fluctuations typically cause fluctuations in the parameters of a network \cite{paulsson}. For example, Fig.\ \ref{fig1}a shows a model of gene expression that includes promoter activation, transcription, translation, and degradation \cite{kepler,kaern05,raser,golding}. In this model, $v_1$ is the rate of translation. It is a function of the number of free ribosomes, an extrinsic variable, and will fluctuate as the number of free ribosomes changes. Extrinsic fluctuations have an average lifetime that is not zero (they are `colored') \cite{rosenfeld05}. We will show that this extrinsic timescale can profoundly affect the system's dynamics and stochastic properties. It can determine the lifetime of protein fluctuations and change mean protein numbers. Extrinsic fluctuations being non-specific can act simultaneously on many parameters of the network. This non-specificity can cause fluctuations to combine constructively or destructively, dramatically altering the network's output. For our simulations, we designed a novel extension of the standard algorithm for simulating intrinsic fluctuations \cite{gillespie76} that includes discontinuous, time-varying parameters and therefore can simulate extrinsic fluctuations with any desired properties (Materials and methods).

\subsection*{Results}
\subsubsection*{Extrinsic fluctuations alter mean protein numbers and intrinsic noise}

\begin{figure}[ht]
\centering
\includegraphics[width=100mm]{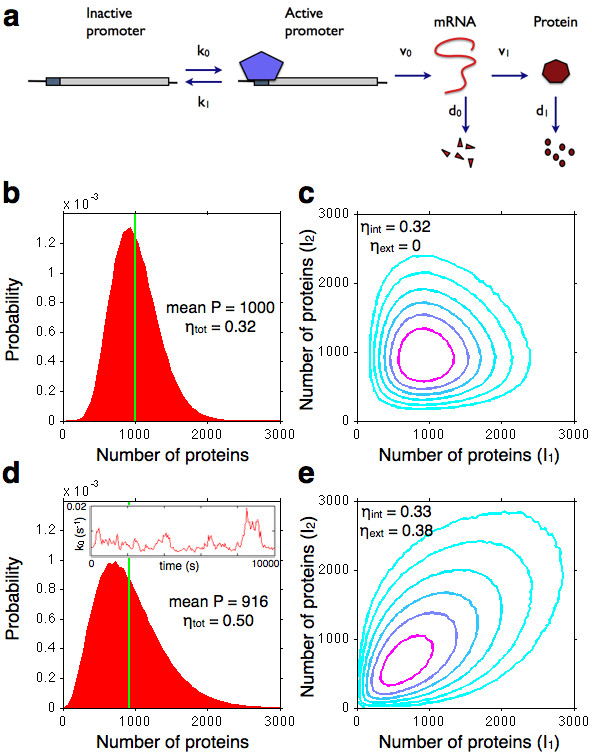}
\caption{Extrinsic noise changes substantially the probability distribution for protein numbers during gene expression. {\bf a} A model of gene expression with two states of the promoter, one active, and able to initiate transcription, and the other inactive. We have shown the binding of RNA polymerase (purple pentagon) driving the transition from active to inactive, but the transition may occur through the binding of transcription factors or changes in the structure of chromatin \cite{blake,raser,golding}. Once active, the promoter can initiate transcription on average every $1/v_0$ seconds and synthesize mRNA which in turn is translated into protein on average every $1/v_1$ seconds. Both protein and mRNA undergo first-order degradation. We use parameters appropriate for {\it Escherichia coli} \cite{golding}: $k_0 = 0.005$ s$^{-1}$, $k_1 = 0.03$ s$^{-1}$, $v_0 = 0.07$ s$^{-1}$, $d_0 = 0.005$ s$^{-1}$, $v_1 = 0.2$ s$^{-1}$, and $d_1 = 0.0004$ s$^{-1}$. The longest intrinsic timescale is then 2500s, the promoter is active approximately 15\% of the time, and the mean steady-state number of proteins is $1000$. {\bf b} A histogram of protein numbers generated by stochastic simulation of {\bf a}. Only intrinsic fluctuations are included. The distribution is slightly skewed with the mode close to the mean, which is shown by the green line. {\bf c} A contour plot of the joint protein probability distribution generated by a two-color experiment for which we simulate two identical copies of the system. {\bf d} A histogram of protein numbers generated from intrinsic fluctions and a fluctuating extrinsic variable: $k_0$, the probability per unit time of the promoter transitioning from the inactive to the active state. The inset shows typical variation of $k_0$. Extrinisic fluctuations are generated by a log-normal stochastic process with an autocorrelation time of approximately $10^3$s: the mean of $k_0$ is unchanged and it has a coefficient of variation of 1. {\bf e} The corresponding joint probability distribution for $I_1$ and $I_2$ in a two-color experiment.}
\label{fig1}
\end{figure}

Extrinsic fluctuations can substantially change the distribution of protein numbers. Fig.\ \ref{fig1}b shows the steady-state distribution of protein numbers for the model of Fig.\ \ref{fig1}a with no extrinsic fluctuations. It is slightly asymmetric and is expected to approximate a gamma distribution \cite{friedman}. Fig.\ \ref{fig1}c shows the corresponding joint probability distribution of $I_1$ and $I_2$.  Although the system is generally described by a probability distribution that includes all the intrinsic variables for the first copy of the network, all the intrinsic variables for the second copy, and all the extrinsic variables, a projection of this distribution onto $I_1$ and $I_2$ is sufficient for calculating noise \cite{elowitz02}. With no extrinsic fluctuations, the distribution spreads parallel to the $I_1$ and $I_2$ axes (Fig.\ \ref{fig1}c): $I_1$ and $I_2$ are independent and have no correlation ($\eta_{\rm ext}=0$). With extrinsic fluctuations, the mode and mean of the distribution of protein numbers can decrease, its variance increases, and there can be a longer tail at high numbers (Fig.\ \ref{fig1}d). Correspondingly, the probability distribution for $I_1$ and $I_2$ spreads along the line $I_1=I_2$: $I_1$ and $I_2$ are now correlated through fluctuations in the extrinsic variable (here, the rate $k_0$: Fig.\ \ref{fig1}d inset). Higher extrinsic noise would cause the distribution to spread along and tighten around the line $I_1=I_2$. Higher intrinsic noise would cause the distribution to expand away from the line.

\begin{figure}[ht]
\centering
\includegraphics[width=160mm]{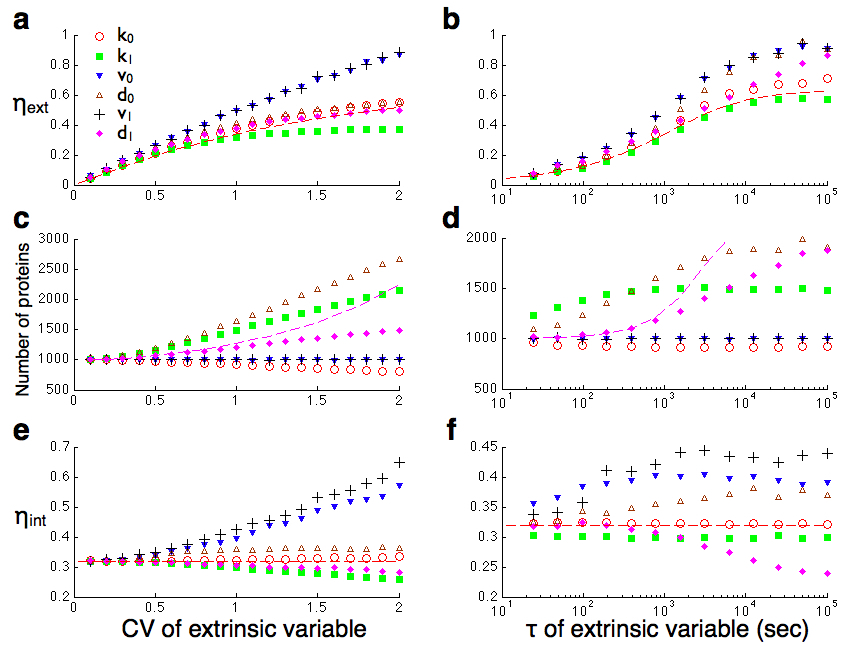}
\caption{The effects of the magnitude of extrinsic fluctuations (the coefficient of variation) and their duration (the autocorrelation time $\tau$) on extrinsic noise ({\bf a} and {\bf b}), mean protein numbers ({\bf c} and {\bf d}), and intrinsic noise ({\bf e} and {\bf f}) for the gene expression of Fig.\ \ref{fig1}a. We simulate fluctuations in either $k_0$ (orange circles), $k_1$ (green squares), $v_0$ (blue triangles), $d_0$ (red triangles), $v_1$ (black crosses), or $d_1$ (purple diamonds). When the coefficient of variation of the extrinsic variable changes, $\tau$ is $10^3$s ({\bf a}, {\bf c}, and {\bf e}). When $\tau$ varies, the coefficient of variation is $1$ ({\bf b}, {\bf d}, and {\bf f}). Each simulation data point is calculated from $10^8$s of simulation. The dashed lines are analytical solutions using the unified colored noise approximation (applied to $d_1$; {\bf c} and {\bf d}) \cite{jung} or Langevin theory (applied to $v_0$; {\bf a}, {\bf b}, {\bf e} and {\bf f}) \cite{vankampen,swain04}. The Langevin approach is suitable for many fluctuating variables, but is a linear approximation and is unable to reproduce the shift in mean protein number. The unified colored noise approximation is non-linear, but is suitable for only one fluctuating variable and cannot be easily applied to the full model (Supplementary information).
}
\label{fig2}
\end{figure}

Changing the properties of extrinsic noise, its source, its magnitude and the typical lifetime of an extrinsic fluctuation ($\tau$), can alter mean protein numbers and the intrinsic noise. The effect of extrinsic fluctuations is determined by both their coefficient of variation and their lifetime. As the coefficient of variation of any parameter in Fig.\ \ref{fig1}a increases, the extrinsic noise in protein numbers increases (Fig.\ \ref{fig2}a). Similarly, as the lifetime of extrinsic fluctuations increases, the extrinsic noise increases: extrinsic fluctuations that are fast compared to intrinsic fluctuations are averaged away and contribute little extrinsic noise (Fig.\ \ref{fig2}b). If the extrinsic fluctuations occur in a parameter that determines the lifetime of fluctuations in protein numbers, such as the protein degradation rate, then the extrinsic timescale mixes with the intrinsic timescales and the mean (and the mode) of the protein distribution can shift (Fig.\ \ref{fig2}c and \ref{fig2}d). Although this change in mean protein numbers implies that extrinsic fluctuations can change intrinsic noise, the change we observe is more than expected (Fig.\ \ref{fig2}e and \ref{fig2}f): if the translation rate, $v_1$, fluctuates, the mean protein number changes little, but there is over a twofold increase in intrinsic noise. To understand this behavior, consider only the measured intrinsic variables and one extrinsic variable, $E$ say, then the system can be described by the probability distribution $P(I_1,I_2,E)$. Changing the properties of the extrinsic variable will change the shape of this three dimensional distribution and consequently its projection onto the $I_1$ and $I_2$ plane (Supplementary information). The intrinsic noise, which is determined by the $I_1$ and $I_2$ projection, can therefore vary with extrinsic fluctuations. We mathematically verified these conclusions for the model of Fig.\ \ref{fig1}a (Fig.\ \ref{fig2} and Supplementary information). 

Our approach also provides a general technique for stochastic sensitivity analysis because we apply fluctuations to parameters of the system \cite{stelling04}. We can therefore determine, for example, the robustness of the concentration of the network output or any other network property to changes in parameter values. Sensitive parameters generate both a high intrinsic and a high extrinsic noise (a high total noise) in the property under investigation. Within our model, we predict that protein levels are most sensitive to fluctuations in the transcription and translation rates, $v_0$ and $v_1$ (Fig.\ \ref{fig2}a and Fig.\ \ref{fig2}e).

\subsubsection*{Extrinsic fluctuations can describe trends in high-throughput measurements}

\begin{figure}[ht]
\centering
\includegraphics[width=150mm]{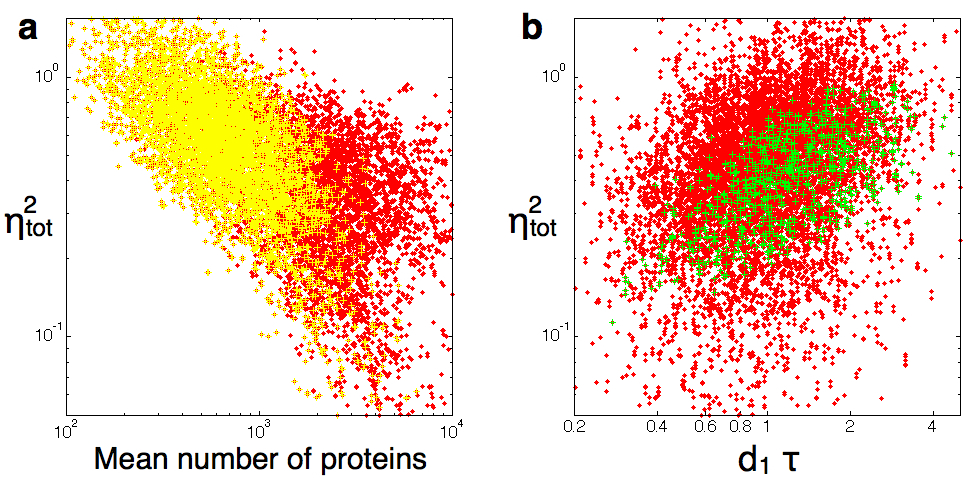}
\caption{Correlations between the the total noise in protein numbers and mean protein numbers and the total noise in protein numbers and the timescale of the extrinsic fluctuations. We randomly generated 10,000 sets of parameters for the model of gene expression in Fig.\ \ref{fig1}a. We added extrinsic fluctuations to one randomly chosen parameter of the model (Materials and methods). Overall, the mean ratio of extrinsic to intrinsic noise in protein numbers is about $2$. {\bf a} The total noise and the mean protein number have a negative correlation of approximately $0.4$ for the entire dataset (red points and yellow crosses). The magnitude of this correlation increases to $0.6$ for parameters where intrinsic noise is at least $50 \%$ of the total noise (yellow crosses). There is also a weak (approximately $0.3$) correlation between the intrinsic and the extrinsic noise. {\bf b} The total noise and the timescale of the extrinsic fluctuations have a  correlation coefficient of approximately $0.35$ for the entire dataset (red points and green crosses). This correlation increases to $0.6$ for parameter sets with extrinsic noise at least $75 \%$ of the total noise (green crosses). }
\label{fig6}
\end{figure}

We can use these properties of extrinsic fluctuations to explain high-throughput measurements of stochasticity. Total noise in protein numbers scales with the inverse square root of mean protein number \cite{bar-even,newman}. Bar-Even et al.\ neglected extrinsic noise to explain this relationship, which is expected for intrinsic noise \cite{bar-even}. We generated many models from the scheme of Fig.\ \ref{fig1}a by log-normally sampling its parameter set (Materials and methods). We simulated these models with substantial extrinsic fluctuations and found that the scaling is still apparent, because extrinsic fluctuations can change mean protein numbers and intrinsic noise (Fig.\ \ref{fig6}a). Our results imply that the high throughout measurements are consistent with other work which shows extrinsic noise to be dominant \cite{elowitz02,raser}. 

Extrinsic fluctuations can cause correlations between the lifetime of protein fluctuations and the extrinsic noise in protein levels, if they have the longest timescale in the system. The lifetime of protein fluctuations will then be determined by the lifetime of the extrinsic fluctuations. With the same simulations, we measure a significant correlation between the timescale of protein fluctuations and total noise (Fig.\ \ref{fig6}b). It arises because many of the models we simulate have extrinsic noise greater than intrinsic noise. The correlation is also evident in Fig.\ \ref{fig2}b. Indeed, time series studies in human cells have shown that the total noise to be correlated with the autocorrelation time of protein levels \cite{sigal}. 

\subsubsection*{Extrinsic fluctuations can affect the performance of genetic networks}

\begin{figure}[ht]
\centering
\includegraphics[width=140mm]{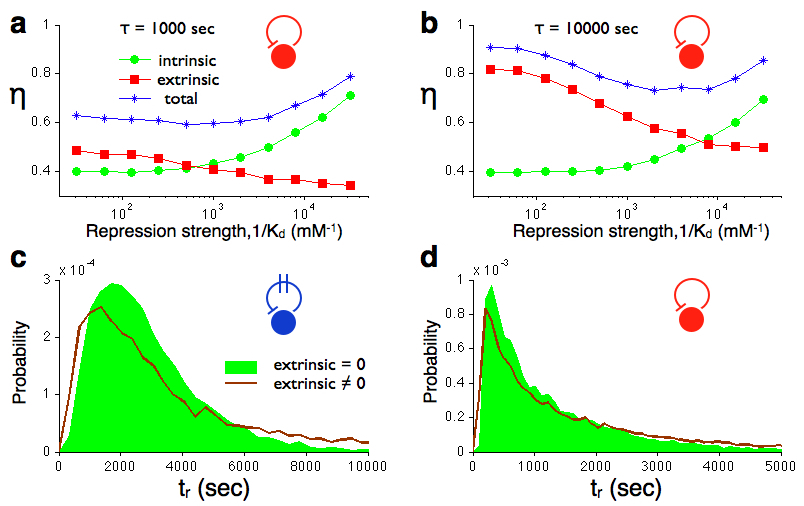}
\caption{Extrinsic fluctuations can enhance the effects of negative auto-regulation on noise and response times. Proteins are repressors and can bind to the inactive promoter state in Fig.\ \ref{fig1}a with a dissociation constant $K_d$. We let $v_0$ have extrinsic fluctuations with a coefficient of variation of 1. Intrinsic noise increases and extrinsic noise decreases as the strength of the feedback increases. The steady-state number of proteins drops from 1000 with no feedback to 300 when the feedback is maximum. {\bf a} For weak extrinsic noise ($\tau= 10^3$s), the total noise mostly increases with feedback strength. {\bf b} For strong extrinsic noise ($\tau= 10^4$s), the total noise mostly decreases with feedback strength. In both cases, there is an optimum $K_d$ for which the intrinsic noise and extrinsic noise are equal and the total noise is minimum. {\bf c} The response time distribution for constitutive expression: measuring from the initiation of transcription, $t_r$ is the time taken to first reach half of the mean steady-state number of proteins. Extrinsic noise ($\tau=10^4$s) decreases the mode of the distribution from 1800s to 1300s, while the mean increases from 2900s to 4700s. {\bf d} The response time distribution for an auto-negative system ($K_d \simeq 60$nM). Extrinsic noise ($\tau=10^4$s) decreases the mode from 300s to 200s. The mean increases from 1300s to 1900s.}
\label{fig3}
\end{figure}

We next considered the effect of extrinsic fluctuations on one of the simplest regulatory networks: a negatively auto-regulated loop. Experiments suggest that total noise is attenuated by negative auto-regulation, at least for a plasmid-borne system \cite{becskei00}. Negative feedback reduces noise by increasing expression when protein numbers are low and decreasing expression when protein numbers are high. It also, however, reduces mean protein numbers. This reduction in protein copy numbers amplifies intrinsic noise and may surpass any attenuating effects (Fig.\ \ref{fig3}a and \ref{fig3}b). Extrinsic noise is mostly independent of protein numbers. It will therefore decrease with the addition of negative feedback. Consequently, the total noise of a constitutive system to which auto-negative feedback is added can either increase or decrease if intrinsic or extrinsic noise is larger (Fig.\ \ref{fig3}a and \ref{fig3}b). Consistently, experiments show a range of auto-repression strength for which noise minimization is optimal, although this observation was attributed to plasmid variation \cite{serrano06}. Our results suggest extrinsic fluctuations in any parameter of the system should create the same effect. We also predict that negative feedback is more likely to evolve as a noise attenuator in systems dominated by extrinsic noise \cite{paulsson,hooshangi06}. Alternatively, intrinsic fluctuations could be reduced by an additional positive feedback loop to maintain high protein copy numbers despite the negative feedback needed to attenuate extrinsic fluctuations. For example, positive and negative feedbacks occur in the GAL regulon in budding yeast and have been shown to reduce fluctuations \cite{ramsey}.

Negative auto-regulation also reduces response times \cite{savageau74,rosenfeld02}. The mean time for an auto-negative system to reach half of its steady-state number of proteins from the initiation of transcription, $t_r$, is reduced by at least a factor of two by negative feedback (Fig.\ \ref{fig3}c and \ref{fig3}d). This reduction occurs because negative feedback decreases the timescales of the system and so shifts the power spectrum of a constitutively expressed gene to higher frequencies \cite{austin}. More intuitively, negative feedback both reduces steady-state protein numbers from constitutive levels and initially allows expression at the higher constitutive rate while the first repressors are synthesized \cite{rosenfeld02}. Stochastic fluctuations can cause significant variation in timing \cite{amir}, and we observe that the probability distribution of $t_r$ is asymmetric and the asymmetry is enhanced by extrinsic noise (Fig.\ \ref{fig3}c and \ref{fig3}d). An extrinsic fluctuation can either aid or inhibit gene expression, and its substantial lifetime ensures that such effects contribute significantly to $t_r$. Despite increasing the mean response time, extrinsic noise enables cells to typically respond faster, irrespective of negative feedback, because the most probable $t_r$ decreases (Fig.\ \ref{fig3}c and \ref{fig3}d). Yet, a population of cells can better `hedges its bets' \cite{kussell} because a greater number will rarely respond: the distribution has a longer tail for high response times.

\subsubsection*{Extrinsic fluctuations can combine destructively and constructively}

\begin{figure}[ht]
\centering
\includegraphics[width=140mm]{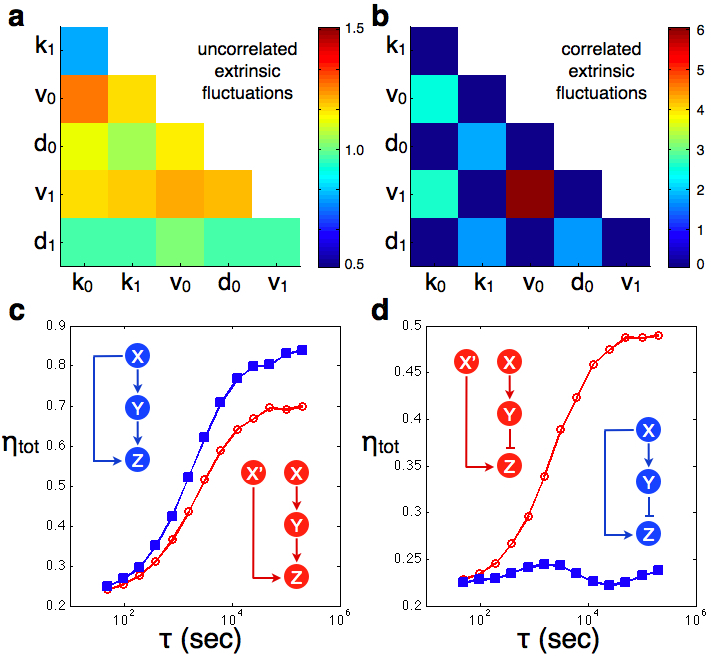}
\caption{Non-specific effects of extrinsic noise: fluctuations can combine constructively or destructively when extrinsic noise is applied to two parameters in Fig.\ \ref{fig1}a. Let $\eta_{\rm ext}^{(i)}$ be the extrinsic noise when the parameter $i$ fluctuates, and let $\eta_{\rm ext}^{(i,j)}$ be the extrinsic noise when parameters $i$ and $j$ fluctuate. {\bf a} The relative extrinsic noise, $(\eta_{\rm ext}^{(i,j)})^2/[ (\eta_{\rm ext}^{(i)})^2 + (\eta_{\rm ext}^{(j)})^2]$, when uncorrelated fluctuations are applied to a pair of parameters. The scale-bar shows the magnitude of the relative extrinsic noise: $(\eta_{\rm ext}^{(i,j)})^2 \simeq (\eta_{\rm ext}^{(i)})^2 + (\eta_{\rm ext}^{(j)})^2$. {\bf b} The relative extrinsic noise when correlated extrinsic fluctuations are applied to a pair of parameters: $ (\eta_{\rm ext}^{(i,j)})^2 \gg (\eta_{\rm ext}^{(i)})^2 + (\eta_{\rm ext}^{(j)})^2$ or $(\eta_{\rm ext}^{(i,j)})^2 \ll (\eta_{\rm ext}^{(i)})^2 + (\eta_{\rm ext}^{(j)})^2$. {\bf c} Total noise in a coherent feedforward loop versus the timescale of extrinsic fluctuations in the number of $X$ proteins. The feedforward amplifies noise: extrinsic fluctuations in $X$ combine constructively with those in $Y$. {\bf d} Total noise in an incoherent feedforward network versus the timescale of extrinsic fluctuations in the number of $X$ proteins. The feedforward loop attenuates noise: extrinsic fluctuations in $X$ combine destructively with those in $Y$. In {\bf c} and {\bf d}, results for the feedforward loop are shown with blue squares and results for the equivalent `open' loop are shown with red circles. In the open loop, $X'$ has the same mean copy number as $X$, but uncorrelated extrinsic fluctuations. The coefficient of variation of the extrinsic fluctuations in $X$ or $X'$ is 1. The two protein inputs $X$ and $Y$ regulate the gene $Z$ through an AND gate. Other types of feedforward or regulation give similar results.}
\label{fig4}
\end{figure}

Extrinsic fluctuations are non-specific: they can act simultaneously on many parameters of a network \cite{pedraza}. We added extrinsic fluctuations to all pairs of parameters in the model of Fig.\ \ref{fig1}a. These fluctuations were either uncorrelated, and generated by individual noise sources, or identical, and generated by the same noise source (Fig.\ \ref{fig4}a and \ref{fig4}b). For uncorrelated extrinsic fluctuations, the extrinsic noise in each parameter combines constructively: the extrinsic noise is approximately the sum of the extrinsic noises generated when each parameter fluctuates alone (Fig.\ \ref{fig4}a). For identical or, more generally, correlated extrinsic fluctuations, the extrinsic noise in each parameter also combines constructively if both parameters affect protein numbers similarly (protein numbers are proportional or inversely proportional to both parameters). Extrinsic fluctuations can be destructive, however, if both parameters have opposing effects on protein numbers (protein numbers are proportional to one parameter and inversely proportional to the other). Fluctuations in the two extrinsic variables then have little effect on extrinsic noise because a fluctuation in the variable that acts to increase protein numbers is counteracted by the same, or a similar, fluctuation in the variable that acts to decrease protein numbers (Fig.\ \ref{fig4}b). A network architecture that channels extrinsic noise into two parameters with opposing effects on protein numbers can therefore attenuate noise, and one that channels extrinsic noise into two parameters with similar effects on protein numbers can be a noise amplifier. We confirmed these results using a Langevin calculation (Supplementary information).

Constructive and destructive extrinsic fluctuations occur in feedforward loops, one of the most common motifs in genetic networks \cite{milo}. Fig.\ \ref{fig4}c and \ref{fig4}d illustrate two feedforwards, where gene $Z$ is activated by genes $X$ and $Y$, and gene $Y$ is either activated by gene X (coherent feedforward) or repressed by gene $X$ (incoherent feedforward) \cite{mangan}. Extrinsic fluctuations in $Y$ can combine constructively or destructively with those in $X$ because the feedforward correlates the fluctuations of $Y$ with the extrinsic fluctuations in $X$. If the timescale of the extrinsic fluctuations is less than intrinsic timescales, however, extrinsic fluctuations are averaged away and such effects are no longer seen \cite{ghosh,hayot}. In the coherent feedforward loop, the extrinsic fluctuations are constructive because $X$ and $Y$ affect gene expression of $Z$ similarly (Fig.\ \ref{fig4}c). In the incoherent feedforward loop, $X$ and $Y$ having opposing affects on gene expression and their extrinsic fluctuations are destructive (Fig.\ \ref{fig4}d). As well as being sign-sensitive delays and accelerators \cite{mangan}, feedforward loops may therefore also have been selected to amplify (coherent) or attenuate (incoherent) extrinsic fluctuations.

\subsection*{Discussion}

Here we have extended the standard stochastic simulation algorithm for simulating intrinsic fluctuations in biochemical networks \cite{gillespie76} to include extrinsic fluctuations (Materials and methods). Although extrinsic fluctuations have been modeled previously \cite{volfson,sigal,geva,austin,cox06,tsimring,scott,wolde}, our approach is more general: we can simulate extrinsic fluctuations with any desired properties; we can vary many parameters with correlated or uncorrelated fluctuations; and we are able to average over intrinsic fluctuations by repeating simulations with the same trajectory of extrinsic variation. In the Supplementary information, we show that time-varying extrinsic fluctuations lead to a generalization of the original definitions of intrinsic and extrinsic noise \cite{swain02}.

Both the magnitude and the timescales of fluctuations are necessary to predict the effects of one stochastic system interacting with another. The mixing of the timescales of the two systems through their interaction can lead to so-called deviant effects \cite{samoilov}, such as a shifting of the mean and asymmetries in the distribution of protein numbers. Extrinsic fluctuations can even decrease intrinsic noise in protein levels. We predict that deviant effects will be common in biochemical networks because these networks typically have substantial extrinsic fluctuations and the timescale of these fluctuations can be the longest timescale in the system. Indeed, such effects are present in high-throughput measurements of cellular variation \cite{bar-even,newman,sigal}. 

We can use our simulation method to investigate the source of extrinsic fluctuations. Interpreting our results as a stochastic sensitivity analysis, we predict that variation in transcription and translation rates to be the most significant sources. Such variation is likely to arise from fluctuations in the numbers of ribosomes and RNA polymerases. Being based on the parameter set of Fig.\ \ref{fig1}a, this prediction is model-specific, but we expect it to hold for other genes in {\it E.\ coli}.

Extrinsic fluctuations can create stochasticity in the output of a network of a magnitude that is substantially different from the magnitude of the extrinsic fluctuations themselves. If correlated, fluctuations in two parameters of a network can combine constructively to create extrinsic noise in the protein output that is many times the extrinsic noise for each parameter fluctuating independently. A different network architecture, however, can cause correlated extrinsic fluctuations to almost entirely negate each other. Both effects are likely to be present in cells.

Extrinsic fluctuations, through their timescales and non-specificity, are thus an important component of the intracellular environment. To function in this environment, biochemical networks are likely to have evolved to control or exploit these fluctuations. Our stochastic simulation algorithm and mathematical analysis should therefore help to quantitatively understand endogenous networks and to design effective synthetic ones.

\subsection*{Materials and methods}

To simulate extrinsic noise, we extend Gillespie's first reaction algorithm \cite{gillespie76} to include discontinuous, time-dependent reaction rates. In the first reaction algorithm, a putative time for each potential reaction in the system is calculated, and the reaction whose putative time is first is implemented. Simulation time is then incremented by this reaction time. Each putative reaction time is calculated from the propensity of the reaction: the probability of the reaction per unit time multiplied by all ways of selecting the reactants \cite{gillespie76}. The propensity, $a(t)$, is a function of time if the probability of the reaction per unit time is not constant.

For a time-varying propensity, we can show (Supplementary information) that the putative reaction time, $\tau$, obeys
\equn{ag}
\int_0^\tau dt \, a(t) = \log(1/r)
\nuqe
where $r$ is a uniform random number between 0 and 1. Eq.\ \ref{ag} is general, but it may be difficult to analytically find $\tau$ for a complex $a(t)$. Consequently, we approximate $a(t)$ by a series of step functions or a piecewise-linear function. If we sample $a(t)$ every $\Delta t$ seconds and use the more accurate piece-wise linear approximation, then $a(t) \simeq a_0 + a_1 t$ for $t$ within a $\Delta t$ interval. Here $a_0$ and $a_1$ are constants defined by the Taylor series of $a(t)$ and will change discontinuously from one $\Delta t$ interval to the next. We can use Eq.\ \ref{ag} to exactly implement discontinuous changes in a propensity (Supplementary information). Briefly, if the next predicted reaction would bring the simulation time into the next $\Delta t$ interval, we do not implement this reaction, but instead change the time-dependent propensity to the new functional form valid for the new $\Delta t$ interval. We set the simulation time to the start of the new $\Delta t$ interval and re-calculate the putative reaction times for all the reactions.

To calculate the putative reaction time within each $\Delta t$ interval when $a(t)= a_0 + a_1 t$, we again use Eq.\ \ref{ag} which implies 
\equn{ag2}
\tau= \frac{a_0}{a_1} \left( \sqrt{1 - \frac{2 a_1}{a_0^2} \log(r)} -1 \right)
\nuqe
where $\tau$ obeys $0 \le \tau \le \Delta t$. If $a_1 < 0$ and $r > \e^\frac{a_0^2}{2a_1}$, then the reaction cannot occur ($\tau= \infty$).

By generating a time series for an extrinsic noise source before running our algorithm, we can then use this time series to change reaction rates appropriately to simulate extrinsic fluctuations. We use the Ornstein-Uhlenbeck process to generate the time series \cite{fox,gillespie92}. This process, $\eta(t)$, has a positive autocorrelation time and is normally distributed. Consequently, when added to a parameter $k$ so that $k \rightarrow k + \eta(t)$, $k$ can become negative. Exponentiating $\eta(t)$, however, and letting  $k \rightarrow k \e^{\eta(t)} / \la \e^{\eta(t)} \ra$ generates a log-normal stochastic process for $k$.  Such a process is suitable for modeling fluctuations in extrinsic variables \cite{rosenfeld05}: $k$ has a fixed mean, a finite auto-correlation time, and is always positive. 

We simulate with the Gibson-Bruck version of the Gillespie method \cite{gibson} using the {\it Facile} network complier and its stochastic simulator \cite{siso}. Both are freely available. All reactions and kinetic rates are included in the Supplementary information. When simulating two copies of the system, we define $\eta^2_{\rm int}= \frac{\la (I_1-I_2)^2 \ra}{2\la I \ra^2}$ and $\eta^2_{\rm ext}= \frac{\la I_1 I_2 \ra - \la I \ra^2}{\la I \ra^2}$ where $I_1$ is the number of proteins for the first copy and $I_2$ the number of proteins for the second copy. We use $\la I_1 \ra = \la I_2 \ra = \la I \ra$ because both copies have the same mean. In our simulations, averages are time averages taken over many times the longest timescale of the system.

For Fig.\ \ref{fig6}, we generated parameter sets for our model of gene expression from log-normal distributions with means given by the parameter values in Fig.\ \ref{fig1}a and a variance in log-space of $20 \%$ of the mean. We choose $k_0$, however, by sampling the probability of the promoter being in the active state from a log-normal distribution with a mean given by the parameters in Fig.\ 1a and with a variance of $70 \%$ of this mean. We let extrinsic fluctuations act on a randomly chosen parameter in each model. These fluctuations have a coefficient of variation of 1. For each model, we sample $\tau$ from a log-normal distribution with a mean of 2500s, the mean protein lifetime, and with a variance of $50\%$ of this mean.

\subsection*{Acknowledgments}
We thank M.\ Chacron, M.\ Elowitz, D.\ Gillespie, and N.\ Rosenfeld for useful conversations. In particular, we thank T.J.\ Perkins for showing us the derivation of Eq.\ \ref{ag}. P.S.S.\ holds a Tier II Canada Research Chair. V.S., J.F.O.\, and P.S.S.\ are supported by N.S.E.R.C.\ and the M.I.T.A.C.S.\ National Centre of Excellence.


\end{document}